\documentclass[aps,prd,twocolumn,nofootinbib,superscriptaddress]{revtex4-2}

\usepackage[T1]{fontenc}        
\usepackage[utf8]{inputenc}     
\usepackage{newtxtext,newtxmath}

\usepackage{amsmath}
\usepackage{graphicx}
\usepackage[caption=false]{subfig}
\usepackage{caption} 
\usepackage{dcolumn}
\usepackage{bm}
\usepackage{enumitem}  
\usepackage[justification=raggedright,singlelinecheck=false]{caption}
\usepackage{xcolor}
\usepackage{ulem}
\usepackage{booktabs}

\begin{document}

\preprint{APS/123-QED}

\title{Relaxation of time-variable neutron-loaded relativistic jets across the photosphere\\ and their GeV–TeV neutrino counterparts}

\author{Kanako Nakama}
\email{kanako.nakama@astr.tohoku.ac.jp}
\affiliation{%
Astronomical Institute, Graduate School of Science, Tohokku University, Aoba, Sendai, 980-8578, Japan 
}%

\author{Kazumi Kashiyama}
 \affiliation{%
Astronomical Institute, Graduate School of Science, Tohokku University, Aoba, Sendai, 980-8578, Japan 
}%

\author{Nobuhiro Shimizu}
\affiliation{%
ICEHAP, Chiba University, Chiba, Japan
}%

\date{\today}

\begin{abstract}
Both observational and theoretical studies indicate that the central engine of a gamma-ray burst (GRB) is intrinsically time-variable, implying jet inhomogeneity. A jet with an inhomogeneous Lorentz factor distribution develops internal shocks both below and above the photosphere, relaxing toward homologous expansion.
Below the photosphere, neutrons, whose mean free paths are much longer than those of charged particles, play an essential role in the dissipation process.
Using neutron-inclusive shell simulations with initial conditions based on the collapsar scenario, we link the statistical inhomogeneity of the jet at the breakout of the progenitor to the dissipation that occurs inside and outside the photosphere, and calculate the GeV–TeV neutrino counterpart originated from inelastic neutron-proton interactions consistently with the prompt gamma-ray emission. 
We find that the peak energy of the GeV–TeV neutrinos is in \( 10-30\,\mathrm{GeV} \) irrespective to the baryon loading factor of the jet, with the high-energy tail extending into the TeV range as the amplitude of the time variability becomes stronger. 
When gamma-ray emission is efficient as in typical GRBs (i.e., the gamma-ray radiation efficiency with respect to the total jet power is $\sim100\%$), the radiative efficiency of GeV–TeV neutrinos remains 0.1-10\%. 
By contrast, when the gamma-ray radiation efficiency is relatively low ($\lesssim 10\%$) for jets where a large fraction of the energy is dissipated below the photosphere, 
the neutrino efficiency can increase up to 20\%. 
This suggests that GRBs with relatively low gamma-ray luminosities, as well as X-ray-rich transients, 
can be promising targets for ongoing and future GeV--TeV neutrino transient searches.

\end{abstract}

\maketitle


\section{\label{sec:level1}INTRODUCTION}

Gamma-ray bursts (GRBs) are among the most luminous transient phenomena in the universe.
Their prompt gamma-ray emissions, characterized by strong variability and non-thermal spectra, indicate that GRBs are powered by ultra-relativistic jets.
The origin of these relativistic jets in long GRBs is thought to be the collapsar scenario~\cite{Woosley1993}, in which the core collapse of a massive star produces a central compact object, either a black hole or a neutron star, surrounded by an accretion disk that powers the jet.
One of the classical radiation mechanisms proposed for GRB prompt emission is the internal-shock model~\cite{ReesMeszaros1994}. Internal shocks, triggered by fluctuations within the jet, reproduce the observed prompt-emission light curves~\cite{KPS1997}.
It has been shown that efficient gamma-ray production is achieved when the jet exhibits short-timescale, large-amplitude fluctuations~\cite{Beloborodov2000,Kobayashi2001}. 

Although the general features of GRB jets and their progenitors are increasingly understood (e.g.,~\cite{KumarZhang2015}), the microphysics of jet formation, acceleration, and dissipation remain a fundamental challenge.
This challenge arises in part because the region near the GRB central engine is extremely hot and dense, making it opaque to electromagnetic observations.
In such a hot and dense environment near the central engine, thermal equilibrium favors a neutron-rich composition, so the jet is expected to carry neutrons in roughly equal numbers to protons~\cite{KohriMineshige2002,Beloborodov2003,LevinsonEichler2003}.
Neutrons that collide inelastically with protons within the photosphere can produce high-energy (GeV–TeV) neutrinos~\cite{MeszarosRees2000,Derishev+1999, BahcallMeszaros2000}, which escape essentially unattenuated. These neutrinos thus provide unique information about the neutron content and its dynamics, allowing us to probe the environment of jet formation and acceleration.

Observational efforts have been made to detect such neutrinos.
In particular, the IceCube Collaboration~\cite{IceCubeCollab2023} constrained canonical jet parameters related to the jet's neutron content, based on the non-detection of neutrinos from GRB 221009A, the brightest GRB ever observed.
To obtain these constraints, they employed a template proposed by Murase et al.~\cite{Murase2022}.  
This template provides a neutrino energy spectrum assuming a constant relative Lorentz factor of \( \Gamma_\mathrm{rel} \sim 2 \) between neutrons and protons, for a given bulk Lorentz factor.

Although the above framework provides a useful benchmark, it does not fully capture the dynamics expected in realistic GRB jets.
In particular, the central engine exhibits temporal variability, which naturally leads to a wide range of relative Lorentz factors, bulk Lorentz factors, and radii for neutron–proton (n-p) collisions.
Fundamentally, jets with different variabilities experience internal dissipation differently, both below and above the photosphere.
In other words, the diversity in jet variability directly results in variations in the energy partition among neutrinos, photons emitted at sub-photospheric and beyond-photospheric scales, and the residual kinetic energy that powers the afterglow emission.
Accurate evaluation of such dissipation requires numerical simulations based on (magneto)hydrodynamics~\cite{RudolphTamborraGottlieb2024,RudolphTamborraGottlieb2025}.  
However, it remains highly challenging to simultaneously capture the defining features of GRB jets and consistently resolve the microphysics of neutron dissipation.

To bridge this gap, we adopt a simplified model to explore the internal relaxation phase of the jet prior to homologous expansion, during which energy is dissipated through collisions between its constituent fluid elements. This approach builds on previous studies~\cite{KPS1997,Beloborodov2000,Kobayashi2001} that modeled such dynamics using shell-based treatments.
We extend these shell-based models by initiating the treatment within the photosphere and newly incorporating relevant sub-photospheric physical processes in which neutrons play an essential role, using initial conditions at the time when the jet emerges from the stellar surface motivated by the collapsar model(e.g., ~\cite{BrombergNakar2011, MizutaIoka2013}).

The paper is organized as follows.
Section~\ref{sec:model} describes our model.
In Section~\ref{sec:result}, we show the main results
of our calculations and the resulting emissions.
Section~\ref{sec:summaryANDdiscussion} is devoted to summary and discussion.

\section{MODEL}\label{sec:model}
We consider jets in the collapsar scenario~\cite{Woosley1993}.
In this scenario, a relativistic jet originates from deep within the progenitor star,
which is a Wolf-Rayet star with a stellar radius of $R_* \sim 10^{11}\,\mathrm{cm}$,
and propagates through its interior, forming the cocoon surrounding the jet.
Under typical parameter choices, the jet is confined by the cocoon, 
preventing significant expansion and acceleration inside the star, 
and instead leading to the formation of a recollimation shock~\cite{Matzner2003,ZhangWoosley2003,BrombergNakar2011}.
After penetrating the stellar envelope, the jet enters a phase of free expansion.

In this study, we begin tracking the jet dynamics from the onset of free expansion, at which point the jet is still located below the photosphere.
We model the irregular jet as a series of relativistic shells, and follow its dynamics beyond the photosphere.
This approach effectively maps the time variability of the engine onto discrete shell injection. 
We assume that the magnetic field has dissipated while the jet drills through the star and is no longer dynamically dominant~\cite{BrombergGranotPiran2015, BrombergTchekhovskoy2016}.

The outline of the method is as follows: equal-mass shells with various baryon loading factors, specific energies, sampled from a log-normal distribution, are injected at time intervals of $\delta t$. Each shell evolves according to scaling relations derived from relativistic hydrodynamics~\cite{Piran1993}.
When the density of each shell becomes sufficiently low, the shells undergo n-p decoupling as well as decoupling from photons.
When a faster shell catches up with a slower one, a collision occurs. The collision is modeled based on relativistic kinematics and hydrodynamics, and the physical quantities of the resulting merged shells are updated at the next time step.

Using this framework, we follow the location and dynamics of internal collisions between shells within the jet, and compute the associated emissions. 
Although the basic framework of our model is the same as that in \cite{Beloborodov2000, Kobayashi2001},
we include the initial conditions based on the collapsar model~\cite{BrombergNakar2011,MizutaIoka2013} and follow the acceleration phase of the shells.
We then model collisions in which significant internal energy remains within the shells, and incorporate sub-photospheric dissipation processes where neutrons play an essential role.

Hereafter, quantities with a prime (${}'$) are in the comoving frame, and quantities with an overbar ($\bar{\, \,}$) are defined in the center-of-momentum frame of the colliding shells. Otherwise, quantities are in the laboratory frame.

\subsection{Initial conditions}
During the propagation inside the progenitor star, the jet cylindrical radius is determined by the force balance at the recollimation shock surface.  
When the jet head reaches the progenitor radius, the jet punches out of the star. 
Before entering the phase of free expansion, however, the jet undergoes a weak expansion, 
during which the Lorentz factor increases by a factor of $\sim 5$~\cite{MizutaIoka2013}.
Consequently, at the onset of free expansion, the initial transverse size of the jet can be estimated as
\begin{align}
R_\mathrm{0} &= \frac{R_\mathrm{j}}{\theta_\mathrm{j}} 
\approx R_\mathrm{j}\,\Gamma_{\rm ini} \nonumber \\
&\sim 2.4 \times 10^9\,{\rm cm} 
\left( \frac{R_\mathrm{j}}{2.4 \times 10^8\,{\rm cm}} \right) 
\left( \frac{\Gamma_{\rm ini}}{10} \right),
\end{align}
where $R_\mathrm{j}$, $\theta_\mathrm{j}$, $\Gamma_\mathrm{ini}$ denote the jet cylindrical radius, the opening angle after the jet break out, and the Lorentz factor at the onset of the free expansion~\cite{BrombergNakar2011,MizutaIoka2013}. 
Here, the opening angle is approximately $\theta_\mathrm{j} \sim 1/\Gamma_\mathrm{ini}$ 
owing to the relativistic beaming effect.
Although $R_\mathrm{j}$, in general, depends on the jet luminosity and injection time, 
for simplicity, we adopt a common value of 
$R_\mathrm{j}\sim 2.4 \times 10^8\,\mathrm{cm},\,\Gamma_{\rm ini}=5\,\mathrm{or}\,10$ for all shells in this work.

From $r = R_\mathrm{0}$, we continuously inject shells with an interval of $\delta t$,
and each shell has a thickness of 
\begin{equation}
   \delta r = c\delta t,
\end{equation}
where $\delta t$ represents the characteristic variable timescale of the jet. 
We denote the total neutron and proton masses carried by shell $\mathrm{i}$ as 
$m_\mathrm{n}^{(i)}$ and $m_\mathrm{p}^{(i)}$, respectively. 
Each shell is initially composed of equal neutron and proton masses, 
i.e., $m_\mathrm{n}^{(i)} = m_\mathrm{p}^{(i)} = m$, yielding a total mass of $2m$.
Each shell has the same initial Lorentz factor, $\Gamma_\mathrm{ini} \sim 10$, but differs in its baryon loading factor, $\eta_{\mathrm{ini,i}}$, and/or internal energy, ${E'_\mathrm{i}}$. These quantities are related by $2\eta_{\mathrm{ini,i}}mc^2 = \Gamma_\mathrm{ini} (2mc^2 + E'_\mathrm{i})$. 
The baryon loading factor is randomly drawn from a log-normal distribution~\cite{Beloborodov2000, IokaNakamura2002}:
\begin{equation}\label{Eq:log-normal_xi}
  P(\xi) = \frac{e^{-\xi^2 / 2} }{\sqrt{2\pi}},
\end{equation}
where
\begin{equation}\label{Eq:log-normal_eta}
  \ln\frac{\eta_{\mathrm{ini}} - 1 }{\eta_0 - 1} = A\xi, 
\end{equation}
 and $\eta_{\mathrm{ini}}$ is restricted to $\eta_{\mathrm{ini}} \ge \Gamma_\mathrm{ini}$, i.e. we set $\Gamma_\mathrm{ini}$ as the lower bound for the baryon loading factor of each shell.

In this model, a time-variable jet is parameterized by $\delta t$, $\eta_0$, $m$, and $A$.
The variability timescale $\delta t$ corresponds to the observed temporal structure of GRB light curves~\cite{KPS1997}, which typically ranges from $\delta t \sim$1 ms to 10 s~\cite{Veres+2023}.
For a set of initial baryon loading factors realized with fixing $\eta_0$ and $m$, the mean baryon loading factor and the isotropic luminosity of the jet can be obtained as 
\begin{equation}
    \left< \eta_\mathrm{ini} \right> \equiv \frac{1}{N} \sum^{N}_{i = 1} \eta_{\mathrm{ini,i}},
\end{equation}
and
\begin{equation}
    L_\mathrm{j,iso} = \frac{\left< \eta_\mathrm{ini} \right> m c^2}{\delta t},
\end{equation}
respectively.
Here N represents the total number of the shells.

In the classical internal shock model, a value of $A \gtrsim 1$ is required to account for 
the high radiation efficiency of GRBs~\cite{Beloborodov2000, Kobayashi2001}.  
Beyond the photosphere, the maximum energy that can be 
released through shell collisions in the entire system is 
$(\langle \eta_\mathrm{ini} \rangle - \Gamma_\mathrm{CM})\,N\,(m_\mathrm{n}^{(i)} + m_\mathrm{p}^{(i)})\,c^2$,  
where $\Gamma_\mathrm{CM}$ is the Lorentz factor of the center of momentum of the system beyond the photosphere~\cite{Beloborodov2000}.  
When $A\ll 1$, $\langle \eta_\mathrm{ini} \rangle \approx \Gamma_\mathrm{CM}$, and the radiation efficiency is low.

\subsection{Jet dynamics without shell collisions}

Let us first describe the jet dynamics and emission processes in the absence of shell collisions.

\subsubsection{Acceleration}
After being injected at  $r = R_0$, the n-p composite shell initially undergoes adiabatic expansion, converting its internal energy into kinetic energy. During this phase, the internal energy decreases as $E_\mathrm{i}' \propto r^{-1}$, while the Lorentz factor increases as
\begin{equation}\label{eq:ad_exp}
\Gamma = \frac{2\eta_{\mathrm{ini}}mc^2}{2mc^2 + E'} \approx \Gamma_\mathrm{ini} \times \left(\frac{r}{R_\mathrm{0}}\right),
\end{equation}
until the shell reaches either the saturation radius~\cite{Piran1993},
\begin{eqnarray}\label{eq:SatRad}
    r_\mathrm{sat} &\approx& R_0 \left(\frac{\eta_{\mathrm{ini}}}{\Gamma_\mathrm{ini}}\right) \notag \\ 
    &\sim& 4.8\times 10^{10}\,\mathrm{cm}\,\left( \frac{R_\mathrm{0}}{2.4 \times 10^{9}~\mathrm{cm}} \right)\left( \frac{\eta_\mathrm{ini}}{200} \right)\left( \frac{\Gamma_\mathrm{ini}}{10} \right)^{-1},
\end{eqnarray}
or the n-p decoupling radius (described in the next subsection).

\subsubsection{Neutron-proton decoupling}
When injected, neutrons and protons are initially coupled via elastic nuclear scattering and are accelerated together according to Eq. (\ref{eq:ad_exp}). 
They decouple when the expected number of elastic scatterings during the dynamical time drops to unity, i.e.,
\begin{equation}
    n_\mathrm{p}' \sigma_\mathrm{np} v'_\mathrm{rel} \frac{r}{c\Gamma}
    \approx
    n_\mathrm{p}' \sigma_\pi \frac{r}{\Gamma}
    = 1,
\end{equation}
where $n_\mathrm{p}' \equiv m_\mathrm{p}^{(i)} /(4\pi r^2 \delta r \Gamma m_\mathrm{p})$ is the comoving proton number density, $\sigma_\mathrm{np}$ is the interaction cross section between protons and neutrons, $m_\mathrm{p}$ denotes the proton mass, and $\sigma_\pi \sim 3\times10^{-26}\,\mathrm{cm^2}$ is a pion production cross section between protons and neutrons\cite{Derishev+1999, BahcallMeszaros2000, KoersGiannios2007}.

For a shell with a sufficiently large baryon loading factor, $\eta_\mathrm{ini} > \Gamma_\mathrm{np}$, the n--p decoupling occurs during the acceleration phase,
where the radius and the Lorentz factor at the decoupling are given by
\begin{align}
r_\mathrm{np} &\sim 5.9\times 10^{10}\,\mathrm{cm} 
\left( \frac{L_\mathrm{j,iso}}{3.5 \times 10^{52}~\mathrm{erg\,s^{-1}}} \right)^{1/3}
\left( \frac{\langle\eta_\mathrm{ini}\rangle}{800} \right)^{-1/3} \notag \\
&\qquad\times
\left( \frac{R_0}{2.4 \times 10^{9}~\mathrm{cm}} \right)^{2/3}
\left( \frac{\Gamma_\mathrm{ini}}{10} \right)^{-2/3},
\label{r_np}
\end{align}
and
\begin{align}
\Gamma_\mathrm{np} 
&\approx \Gamma_\mathrm{ini} \frac{r_\mathrm{np}}{R_0}
\sim 250\,
\left( \frac{L_\mathrm{j,iso}}{3.5 \times 10^{52}~\mathrm{erg\,s^{-1}}} \right)^{1/3}
\left( \frac{\langle\eta_\mathrm{ini}\rangle}{800} \right)^{-1/3} \notag \\
&\qquad\times
\left( \frac{R_0}{2.4 \times 10^{9}~\mathrm{cm}} \right)^{-1/3}
\left( \frac{\Gamma_\mathrm{ini}}{10} \right)^{1/3},
\label{Gamma_np}
\end{align}
respectively. 

In this case, the neutron shell enters a coasting phase after decoupling, while the proton shell continues to accelerate.
If the baryon loading parameter additionally satisfies the following condition:
\begin{equation}\label{dec_cond}
\eta_\mathrm{ini} + (\eta_\mathrm{ini} - \Gamma_\mathrm{np}) > 2 \Gamma_\mathrm{rel, dec} \Gamma_\mathrm{np},
\end{equation}
the final scatterings between neutrons and protons become inelastic, leading to neutrino emission~\cite{BahcallMeszaros2000}.
Here $\Gamma_\mathrm{rel,dec}\approx 1.4$ is the typical relative Lorentz factor between neutrons and protons at the decoupling
\footnote{
We simplify the treatment by assuming that all shells follow the steady-state scaling law~\cite{BahcallMeszaros2000}. In general, however, the acceleration can deviate from this scaling due to the velocity dependence of the cross section prior to the final scattering~\cite{RossiBeloborodov2006, KoersGiannios2007}.
}. 
The energy carried away by the neutrinos can be estimated as 
\begin{equation}
    {\cal E}_\mathrm{dec, \nu} = \tau_\mathrm{dec} \kappa_\mathrm{dec} \Gamma_\mathrm{rel, dec} \Gamma_\mathrm{np} mc^2,
\label{eq:E_decNu}
\end{equation}
where $\tau_{\mathrm{dec}}$ is the optical depth for this inelastic scattering 
and $\kappa_{\mathrm{dec}}$ is the corresponding inelasticity for this neutrino production. 
In this work, we adopt $\tau_{\mathrm{dec}} \sim 0.1$ 
and $\kappa_{\mathrm{dec}} \sim 0.03$~\cite{BahcallMeszaros2000,KoersGiannios2007}.

\subsubsection{Photospheric emission}

Radiation decouples from the proton shell when the expected number of Thomson scatterings within the dynamical time drops to unity, i.e., when \(\tau_\mathrm{T,dyn}\equiv n_\mathrm{p}' \sigma_\mathrm{T} r/\Gamma = 1 \), where $\sigma_\mathrm{T} \sim 6.65\times10^{-25}\,\mathrm{cm^2}$ is the Thomson cross section.  

If the proton shell has a sufficiently large baryon loading factor $\eta_\mathrm{ini} > \Gamma_\mathrm{ph}$, the photosphere is reached during the acceleration phase, where the radius and the Lorentz factor at the photosphere are given by 
\begin{align}
r_\mathrm{ph} 
&\sim 1.6 \times 10^{11}~\mathrm{cm}
\left( \frac{L_\mathrm{j,iso}}{3.5 \times 10^{52}~\mathrm{erg\,s^{-1}}} \right)^{1/3}
\left( \frac{\langle \eta_\mathrm{ini} \rangle}{800} \right)^{-1/3} \notag \\[6pt]
&\quad \times
\left( \frac{R_{0}}{2.4 \times 10^{9}~\mathrm{cm}} \right)^{2/3}
\left( \frac{\Gamma_\mathrm{ini}}{10} \right)^{-2/3},
\label{r_ph}
\end{align}

\begin{align}
    \Gamma_\mathrm{ph} 
    \sim 670 \left( \frac{L_\mathrm{j,iso}}{3.5 \times 10^{52}~\mathrm{erg\,s^{-1}}} \right)^{1/3}
\left( \frac{\langle\eta_\mathrm{ini}\rangle}{ 800} \right)^{-1/3}\\
\times\left( \frac{R_\mathrm{0}}{2.4 \times 10^{9}~\mathrm{cm}} \right)^{-1/3}
\left( \frac{\Gamma_\mathrm{ini}}{10} \right)^{1/3},
\label{Gamma_ph}
\end{align}
respectively. In this case, the remaining internal energy of 
\begin{equation}\label{eq:Ephgam}
    {\cal E}_\mathrm{ph, \gamma} \approx (\eta_\mathrm{ini} - \Gamma_\mathrm{ph}) m c^2. 
\end{equation}
is radiated as photospheric emission. 
After crossing the photosphere, the shell is assumed to maintain the Lorentz factor acquired at the photospheric radius.

\subsubsection{Neutron Decay}

Neutrons eventually decay into protons via $\beta$ decay.  
This occurs at a characteristic radius of  
\begin{equation}
    r_\beta \sim 2.7 \times 10^{15}\,\mathrm{cm} \left(\frac{\Gamma}{100}\right).
\end{equation}
In this study, we stop tracking the evolution of neutron shells well before they reach this radius, and therefore neglect the effects of neutron decay.

\subsection{Internal shocks}
When the jet breaks out of the star, it possesses a non-uniform distribution of internal energy.
As the jet gradually converts its internal energy into kinetic energy, this non-uniformity is translated into velocity differences within the jet.
Such a jet undergoes an internal relaxation process until it reaches homologous expansion.
This relaxation proceeds through internal collisions among the fluid elements of the jet, namely between the shells.

A shell collision results in forming shock only when both the rapid shell and the slow shell include protons; a neutron only shell, by construction, is in a state after n-p decoupling.
First, the modeling of collisions accompanied by shocks is presented in this subsection.

 \subsubsection{Energy and momentum conservation}\label{ShellColWS}
We begin by describing the general conservation of energy and momentum before and after collisions with shocks, and then present our modeling approach, which depends on the shell composition and whether the collision occurs below or beyond the photosphere.
Hereafter, quantities with the subscripts $\mathrm{s}$ and $\mathrm{r}$ represent values of the slow and rapid shells, respectively, and those with the subscripts $\mathrm{bf}$ and $\mathrm{af}$ represent values before and after the collision, respectively.
The energy and momentum conservation before and after the collision can be described as 
\begin{align}\label{eq:ene_con}
    (M'_\mathrm{s,af}\bar{\Gamma}_\mathrm{s, af}
    + M'_\mathrm{r,af}\bar{\Gamma}_\mathrm{r, af}) \Gamma_\mathrm{CD}c^2 + {\cal E} \notag
  \\= M'_\mathrm{s, bf}\Gamma_\mathrm{s, bf}c^2 + M'_\mathrm{r, bf}\Gamma_\mathrm{r, bf}c^2,
\end{align}
\begin{equation}
  M'_\mathrm{s, af}\bar{\Gamma}_\mathrm{s, af}\bar{\beta}_\mathrm{s, af}c
  = M'_\mathrm{r, af}\bar{\Gamma}_\mathrm{r, af}\bar{\beta}_\mathrm{r, af}c,
\end{equation}
respectively.
Here
\begin{equation}
    M'_{\mathrm{i},\mathrm{j}} = m_\mathrm{i} + E'_\mathrm{i, j}/c^2 \quad (\mathrm{i} = \mathrm{r\,or\,s}, \,\,\, \mathrm{j} = \mathrm{bf \, or \, af}),
\end{equation}
with $m_\mathrm{i}$ and $E'_\mathrm{i, j}$ being the mass (we assume that the baryon number of each shell is conserved before and after the collision~\cite{Kobayashi2001})
and internal energy of the shell, 
${\cal E}$ represents the energy dissipated and carried away from the system, and
\begin{equation}
  \Gamma_\mathrm{CD} = \frac{M'_\mathrm{r, bf}\Gamma_\mathrm{r, bf} + M'_\mathrm{s, bf}\Gamma_\mathrm{s, bf}}{
  \sqrt{{M'_\mathrm{r, bf}}^2 + {M'_\mathrm{s, bf}}^2 + 2M'_\mathrm{r, bf}M'_\mathrm{s, bf}\Gamma_\mathrm{r, bf}\Gamma_\mathrm{s, bf}(1 - \beta_\mathrm{r, bf}\beta_\mathrm{s, bf})}}
\end{equation}
is the Lorentz factor of the merged shell~\cite{KPS1997}, which yields
\begin{align}
  \Gamma_\mathrm{r, af} &= \bar{\Gamma}_\mathrm{r, af}\Gamma_\mathrm{CD} - \sqrt{\bar{\Gamma}_\mathrm{r, af}^2 - 1} \sqrt{\Gamma_\mathrm{CD}^2 - 1}, \\
  \Gamma_\mathrm{s, af} &= \bar{\Gamma}_\mathrm{s, af}\Gamma_\mathrm{CD} + \sqrt{\bar{\Gamma}_\mathrm{s, af}^2 - 1} \sqrt{\Gamma_\mathrm{CD}^2 - 1} \label{eq:Lorentz_trans_gam}.
\end{align}
We assume that, in the cente-of-momentum frame, internal energies from the pre-collision stage remain within the respective shells:
\begin{equation}
  E'_{\mathrm{i,af}} = E'_{\mathrm{i,bf}} \frac{\bar{\Gamma}_\mathrm{i,bf}}{\bar{\Gamma}_\mathrm{i,af}} \quad (\mathrm{i} = \mathrm{r}, \mathrm{s}).
\end{equation}

By fixing a parameter  $\cal E$, 
which depend on the dissipation processes and the properties of the shock, as shown in the next subsections, 
we determine ($\Gamma_\mathrm{s,af}$, $\Gamma_\mathrm{r,af}$) for given shell properties before the collision ($\Gamma_\mathrm{s,bf}$, $M'_\mathrm{s,bf}$, $\Gamma_\mathrm{r,bf}$, $M'_\mathrm{r,bf}$), by solving Eqs. (\ref{eq:ene_con}–\ref{eq:Lorentz_trans_gam}).
We note that our calculations reduce to those of~\cite{Kobayashi2001} when $E'_{\mathrm{i,bf}} = 0$, i.e., for shocks occurring beyond the photospheres of the shells.

In the post-collision state, if \( E'_{\mathrm{i,af}} > 0 \), shell i (i = r, s) accelerates after the collision.  
In this case, instead of Eqs.~(\ref{eq:ad_exp}) and (\ref{eq:SatRad}), we have
\begin{equation}
    \Gamma \approx \Gamma_{\mathrm{i,af}}\times
    \left( \frac{r}{R_{\mathrm{col}}} \right),
    \quad\quad
    \left( \Gamma < \Gamma_{\mathrm{i,af}} 
    + \frac{E'_{\mathrm{i,af}}}{m_\mathrm{i}c^2} \Gamma_{\mathrm{i,af}} \right),
\end{equation}
where $R_{\mathrm{col}}$ is the radius at which the collision occurs.

\subsubsection{Dissipation and emission}\label{sec:EMISSION}
The quantity ${\cal E}$ in Eq.~(\ref{eq:ene_con}) includes the energy carried away as emission.
In this section, we first estimate the energy carried away by neutrinos produced at shocks, and then by photons produced at shocks.

Neutrinos can be produced when neutrons, owing to their much longer mean free path, traverse the shock front and undergo inelastic collisions with protons~\cite{Beloborodov2017}.
The neutrino energy in the observer frame is approximated as
\begin{equation}
{\cal E_\mathrm{\nu}} \;=\; {\cal E_\mathrm{\nu,r}} + {\cal E_\mathrm{\nu,s}},
\label{eq:Enu_compact}
\end{equation}
where
\begin{equation}
{\cal E_\mathrm{\nu,i}} \;=\;\Gamma_{\rm CD}\,\kappa\,\bigl(\bar{\Gamma}_{\rm i,{\rm bf}}-1\bigr) \, m^{(i)}_\mathrm{n}c^{2}\,\bigl(1-e^{-\tau_\mathrm{np}}\bigr),
\label{eq:Enu_eachShell}
\end{equation}
and 
\[
\tau_\mathrm{np} = n'_\mathrm{p}  \Gamma \sigma_\mathrm{np}  \delta r,\,
\]
with $\sigma_\mathrm{np}\sim 4\times 10^{-26}\,\mathrm{cm^2}$.
Here $\kappa$ is the effective inelasticity due to neutrinos, which we model using the result of \textsc{Geant4} (section~\ref{Geant4_explain}).
For collisions that are too mild to produce pions (i.e., $\Gamma_\mathrm{rel} < 1.4$), $\kappa$ is essentially zero. 
As $\Gamma_\mathrm{rel}$ increases, $\kappa$ rises and reaches values of $\sim 0.1$–$0.2$ for relativistic collisions.
At very high $\Gamma_\mathrm{rel}$ ($\gtrsim 10^{2}$–$10^{3}$), $\kappa$ saturates at around $\sim 0.2$.

Whether photons produced at the shock can be observed is determined by the Thomson optical depth, defined as
\begin{equation}
    \tau_\mathrm{T} = n'_\mathrm{p}\,\Gamma\,\sigma_\mathrm{T}\,\delta r .
    \label{eq:tauT}
\end{equation}
When $\tau_\mathrm{T} > 1$, the shock is a radiation mediated shock (RMS),
where a large portion of the upstream bulk kinetic energy is dissipated into radiation 
trapped behind the shock. 
The deceleration of the upstream flow is then caused by photons emitted from 
the immediate post-shock region~\cite{LevinsonNakar2020}; the particle acceleration is basically suppressed (but see \cite{Derishev+2003, KashiyamaMuraseMeszaros2013}).
In this case, photons originating from neutral pions produced by inelastic neutron scatterings are also not observable.
When $\tau_\mathrm{T} < 1$, the shock is collisionless and charged particles are accelerated,
leading to non-thermal radiation, in particular electron synchrotron radiation~\cite{ReesMeszaros1994, MuraseIoka2013}. 
In addition, photons originating from neutral pions produced by inelastic neutron scatterings can also be observed.
Thus, the photon energy emitted at the shock can be modeled as follows.
\begin{equation}\label{eq:photonFROMnpcol}
{\cal E_\mathrm{\gamma}} =
\begin{cases}
0, &(\tau_\mathrm{T} \ge1), \\
    {\cal E}_\mathrm{\nu} + \epsilon_\mathrm{e}[(\bar\Gamma_\mathrm{r,bf} - 1) + (\bar\Gamma_\mathrm{s,bf} - 1)]mc^2\Gamma_\mathrm{CD}, \quad&(\tau_\mathrm{T} <1). 
\end{cases}
\end{equation}
In this study, we adopt $\epsilon_\mathrm{e} \sim 1/3$.

For the shock without neutrons, we set \( \mathcal{E} = \mathcal{E}_\mathrm{\gamma} \)  in Eqs.~(\ref{eq:ene_con}–\ref{eq:Lorentz_trans_gam}), 
to determine the final state of each shell.

\subsubsection{Neutrons at the shock}\label{sec:TREATMENTofNEUTRONS}
The subsequent dynamics of the neutrons included in the shell forming the internal shock depend on $\tau_\mathrm{np}$.
If \( \tau_\mathrm{np} > 1 \),  
the neutrons are predominantly dragged by the protons originally belonging to the same shell, in the zeroth-order approximation,   
therefore, for i (= r, s) shell containing neutrons in a \( \tau_\mathrm{np} > 1 \) shock, its mass is taken to be  
\( m_\mathrm{i} = 2m \), and we set \( \mathcal{E} = \mathcal{E}_\mathrm{\nu} +\mathcal{E}_\mathrm{\gamma} \)  in Eqs.~(\ref{eq:ene_con}–\ref{eq:Lorentz_trans_gam}), 
to determine the final state of each shell.

On the other hand, 
if \( \tau_\mathrm{np} \le 1 \),  
the neutrons are not dragged by the protons originally belonging to the same shell, and to be decoupled.
Therefore, for i (= r, s) shell containing neutrons in a \( \tau_\mathrm{np} \le 1 \) shock, its mass is taken to be  \( m_\mathrm{i} = m \),
and treat it as independent neutron shell and proton shell components in the final state.  
We approximate that the colliding n–p particles equally share their energy in the observer frame;
\footnote{We confirmed the validity of this treatment by checking the final-state distribution of the n-p collision
 using \textsc{Geant4}, and found that it is reasonably accurate as an average characterization.} 
the post-collision specific energy of the neutrons originally belonging to shell \( \mathrm{i} \) (\( \mathrm{i} = \mathrm{r, s} \))  
is approximated as  
$0.5\left( \Gamma_\mathrm{i,bf} + \Gamma_\mathrm{CD} - {\cal{E}_\mathrm{\nu,i}}/\tau_\mathrm{np}mc^2 \right).$
The Lorentz factor of the newly independent neutron shell in the post-collision state,  
\( \Gamma_{\mathrm{n}(i),\mathrm{af}} \) (\( \mathrm{i} = \mathrm{r, s} \)),  
is defined as the average over neutrons that have undergone collisions and those that have not;
\begin{align}\label{Eq:AnotherShockInducedNPdec_n}
    \Gamma_\mathrm{n(i),af} \approx (1-\tau_\mathrm{np})\Gamma_\mathrm{i,bf} + \frac{1}{2}\tau_\mathrm{np}\left(\Gamma_\mathrm{i,bf}+\Gamma_\mathrm{CD}-\frac{\cal{E}_\mathrm{\nu,i}}{\tau_\mathrm{np} mc^2}\right) .
\end{align}
As a result of collisions with neutrons with $\tau_\mathrm{np}\leq 1$, the Lorentz factor of the contact discontinuity of the proton shock is also modified as 

\begin{equation}\label{Eq:AnotherShockInducedNPdec_CD}
 \Gamma_\mathrm{CD,af} \approx 
\Gamma_\mathrm{CD}
 + 
  \frac{1}{4}\frac{\tau_\mathrm{np}} {m}\left(    \sum_{i=r,s}  m^{(i)}_\mathrm{n} (\Gamma_\mathrm{i,bf}-\Gamma_\mathrm{CD}) -\frac{\cal{E}_\mathrm{\nu}}{\tau_\mathrm{np}c^2}\right).
\end{equation}
The final state of the two proton shells is obtained by replacing 
\(\Gamma_\mathrm{CD} \rightarrow \Gamma_\mathrm{CD,af}\), and
\begin{align}
    \mathcal{E} =  \mathcal{E}_\mathrm{\nu} +\mathcal{E}_\mathrm{\gamma} 
        + \mathcal{E}_\mathrm{N}, 
\end{align}
to solve Eq.~(\ref{eq:ene_con}–\ref{eq:Lorentz_trans_gam}).
Here,
\begin{align}
    \mathcal{E}_\mathrm{N} = 
         \sum_{i=r,s} m^{(i)}_\mathrm{n} 
         \left( \Gamma_{\mathrm{n(i),af}} - \Gamma_{\mathrm{n(i),bf}} \right)c^2 ,
\end{align}
where the term
$m^{(i)}_\mathrm{n} \left( \Gamma_{\mathrm{n(i),af}} - \Gamma_{\mathrm{n(i),bf}} \right)c^2$
denotes the energy gained by neutrons in shell $\mathrm{i}$ through n-p collisions.

\subsection{Neutron streaming after decoupling}\label{ColWithoutS}
As noted earlier, a neutron-only shell represents the state after decoupling.  
When such a neutron shell undergoes a collision, its evolution cannot be described by the kinematic Eqs.~(\ref{eq:ene_con}–\ref{eq:Lorentz_trans_gam}),  
which are based on hydrodynamics, and a different treatment is required.
The approximation that the neutron shell can be treated as a single shell in the post-collision state breaks down because neutron scattering is a stochastic process.  
When $\tau_\mathrm{np} < 1$, only a fraction of the neutrons undergo collisions and are decelerated (or accelerated).
For simplisicity, we enforce shell number conservation even in the post-collision state, and model neutron streaming after decoupling so that energy conservation holds in the observer frame.

First, the energy carried by neutrinos from inelastic neutron-proton or neutron-neutron collisions can be estimated as follows;
\begin{align}
    {\cal{E}}_\mathrm{\nu,Ncol} \!&\approx 
     \Gamma_\mathrm{s,bf}(\Gamma_\mathrm{rel}-1)\kappa mc^2\left(1-e^{-{\tau_\mathrm{np,tot}}}\right)
\end{align}
where 
\[
\Gamma_\mathrm{rel} \approx \frac{1}{2}\left( \frac{\Gamma_\mathrm{r,bf}}{\Gamma_\mathrm{s,bf}} + \frac{\Gamma_\mathrm{s,bf}}{\Gamma_\mathrm{r,bf}} \right),
\]
and
\[
\tau_\mathrm{np,tot}=-\tfrac{\tau_\mathrm{np}}{m}[\sum_{\mathrm{i=r, s}} m^{(i)}_\mathrm{n}+m^{(i)}_\mathrm{p} - m]\]
is the total optical depth of inelastic neutron-proton or/and neutron-neutron collisions
~\cite{Murase2022}.
As in Eq.~(\ref{eq:photonFROMnpcol}) if the n-p collision occur beyond the photosphere, photons from pion decay can reach observer, although it is sub-dominant due to the smallness of $\tau_\mathrm{np}$ beyond the photosphere;
\[
{\cal{E}}_\mathrm{\gamma,Ncol}=
\begin{cases}
    0, &(\tau_\mathrm{T,dyn}\ge1),\\
    {\cal{E}}_\mathrm{\nu} &(\tau_\mathrm{T,dyn}<1).\\
\end{cases}
\]

As in Eq.~(\ref{Eq:AnotherShockInducedNPdec_n}), (\ref{Eq:AnotherShockInducedNPdec_CD}), we approximate that the colliding n–p particles equally share their energy in the observer frame. 
In this case, the Lorentz factors of $\mathrm{i}\,(=\mathrm{r, s})$ shell in the post-collision state is given by
\begin{align}
    \Gamma_\mathrm{i,af} &= \left(1-\frac{m}{m^{(i)}_\mathrm{n}+m^{(i)}_\mathrm{p}}\tau_\mathrm{np}\right)\Gamma_\mathrm{i,bf} \notag\\&+ \frac{1}{2}\frac{m} {m^{(i)}_\mathrm{n}+m^{(i)}_\mathrm{p}}\tau_\mathrm{np}(\Gamma_\mathrm{n,bf}+\Gamma_\mathrm{i,bf})-\frac{1}{2}\frac{\mathcal{E}_{\nu,\mathrm{Ncol}} + \mathcal{E}_{\gamma,\mathrm{Ncol}} }{\left(m^{(i)}_\mathrm{n}+m^{(i)}_\mathrm{p}\right)c^2}\notag.
\end{align}

\subsection{Neutrino spectrum calculation}\label{Geant4_explain}
The neutrinos produced by n-p collisions are calculated using the Monte Carlo simulation toolkit \textsc{Geant4}~\cite{Geant_2003, Geant_2006, Geant_2016}.
In our model, scattering due to the nuclear force occurs not only between neutrons and protons 
but also between neutrons themselves. 
In this work, we approximated the neutrino spectrum from neutron-neutron scattering 
by substituting it with that from neutron-proton scattering.
Our simulation was based on the example file \textit{Hadr03}. 
We placed hydrogen gas as the target and inject neutron beams. 
For each relative kinetic energy, we simulated n-p collisions and extracted 
the momentum distribution of the resulting charged pions in the proton rest frame.
Using the neutrino spectra from pion decay, we estimate the energy of the resulting neutrinos and determine the effective inelasticity due to neutrinos \( \kappa \).  
Finally, the obtained neutrino spectra are Lorentz-transformed from the target rest frame to the observer frame.

During the jet dynamics considered in this work, very high-energy (TeV–PeV) neutrinos can be produced by the dissipation of shock-accelerated hadrons both inside and beyond the photosphere~\cite{Derishev+2003, KashiyamaMuraseMeszaros2013, Dermer2003,Murase2008}.
In this study, we do not consider these neutrinos to avoid uncertainties in the hadronic acceleration process. 
Neglecting them does not significantly affect the shell dynamics and the relaxation of the jet.
Although we assume that the internal energy generated in collisionless shocks is radiated via synchrotron emission, a portion of this energy might in reality be carried away by neutrinos.

\section{RESULT}\label{sec:result}

\begin{figure}[t]
\captionsetup{justification=raggedright, singlelinecheck=false}
\includegraphics[width=0.48\textwidth]{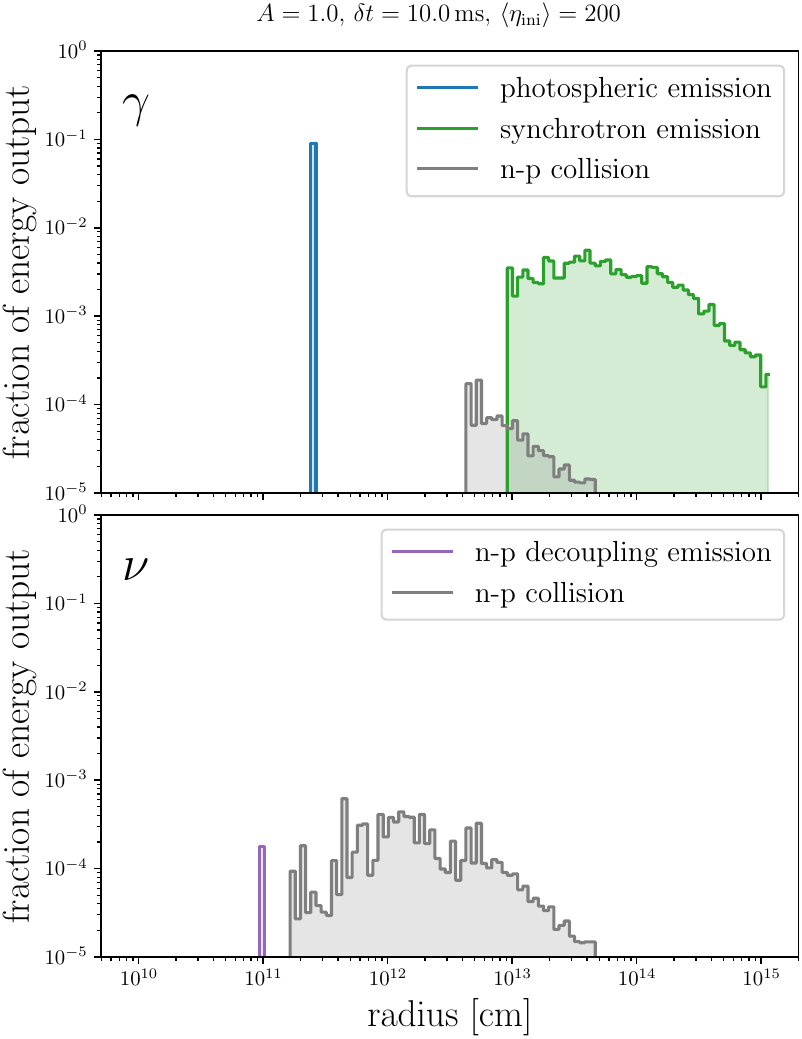}
  \caption{Energy dissipation by photon radiation (upper panel) and neutrino radiation (lower panel) and their dependence on the radius for a time-variable neutron-loaded jet with \( A = 1 \), \( \delta t = 10\,\mathrm{ms} \),  \( \left< \eta_\mathrm{ini} \right> = 200 \), and \( L_\mathrm{j, iso}  = 3.5\times 10^{52}\,\mathrm{erg\,s^{-1}}\). 
  }
  \label{fig:FiducialEneOut}
\end{figure}

\begin{table}[t]
\caption{
Summary of the input parameters and the resulting quantities obtained in this study. 
The first three columns list the input parameters. 
The fourth and fifth columns show the fractions of the injected energy dissipated as photons and neutrinos, respectively (the same quantities as in Figure~\ref{fig:efficiency_all}). 
The sixth and seventh columns give the average Lorentz factors of the proton and neutron shells at the end of the calculation ($t \sim 2\times10^{4}\,\mathrm{s}$ in the central-engine rest frame).
}
\label{tab:energy-results}
\centering
\begin{ruledtabular}
\begin{tabular}{rrr|llll}
 $\langle \eta_{\mathrm{ini}} \rangle$ & $A$ & $\delta t$ [ms] & $E_\gamma / E_\mathrm{tot}$ & $E_\nu / E_\mathrm{tot}$ & $\langle \eta \rangle_\mathrm{p}$ & $\langle \eta \rangle_\mathrm{n}$ \\
\hline
 800 & 0.5 & 1   & $4.6\times10^{-1}$ & $5.7\times10^{-4}$  & 620 & 250 \\
 800 & 0.5 & 10  & $4.4\times10^{-1}$ & $5.7\times10^{-4}$  & 630 & 250 \\
 800 & 0.5 & 100 & $4.7\times10^{-1}$ & $5.1\times10^{-4}$  & 620 & 250 \\
 800 & 1.0 & 1   & $5.9\times10^{-1}$ & $7.4\times10^{-4}$  & 410 & 230 \\
 800 & 1.0 & 10  & $5.9\times10^{-1}$ & $5.4\times10^{-4}$  & 420 & 220 \\
 800 & 1.0 & 100 & $5.8\times10^{-1}$ & $3.6\times10^{-4}$  & 460 & 230 \\
 800 & 2.0 & 1   & $7.2\times10^{-1}$ & $6.7\times10^{-3}$  & 210 & 220 \\
 800 & 2.0 & 10  & $7.8\times10^{-1}$ & $2.8\times10^{-3}$  & 180 & 170 \\
 800 & 2.0 & 100 & $7.9\times10^{-1}$ & $1.0\times10^{-3}$  & 190 & 150 \\
 200 & 0.5 & 1   & $5.1\times10^{-2}$ & $2.0\times10^{-3}$  & 180 & 200 \\
 200 & 0.5 & 10  & $6.1\times10^{-2}$ & $5.7\times10^{-4}$  & 190 & 200 \\
 200 & 0.5 & 100 & $4.6\times10^{-2}$ & $1.2\times10^{-4}$  & 200 & 200 \\
 200 & 1.0 & 1   & $1.4\times10^{-1}$ & $1.7\times10^{-2}$  & 160 & 180 \\
 200 & 1.0 & 10  & $2.2\times10^{-1}$ & $8.9\times10^{-3}$  & 150 & 170 \\
 200 & 1.0 & 100 & $2.3\times10^{-1}$ & $3.2\times10^{-3}$  & 160 & 160 \\
 200 & 2.0 & 1   & $1.9\times10^{-1}$ & $4.3\times10^{-2}$  & 140 & 170 \\
 200 & 2.0 & 10  & $3.6\times10^{-1}$ & $2.6\times10^{-2}$  & 120 & 140 \\
 200 & 2.0 & 100 & $3.7\times10^{-1}$ & $1.3\times10^{-2}$  & 120 & 140 \\
  50 & 0.5 & 1   & $4.7\times10^{-3}$ & $8.1\times10^{-3}$  & 49  & 49  \\
  50 & 0.5 & 10  & $1.3\times10^{-2}$ & $3.6\times10^{-3}$  & 47  & 48  \\
  50 & 0.5 & 100 & $1.3\times10^{-2}$ & $4.3\times10^{-3}$  & 48  & 52  \\
  50 & 1.0 & 1   & $4.4\times10^{-3}$ & $1.4\times10^{-1}$  & 42  & 42  \\
  50 & 1.0 & 10  & $1.3\times10^{-2}$ & $9.1\times10^{-2}$  & 42  & 43  \\
  50 & 1.0 & 100 & $2.4\times10^{-2}$ & $3.6\times10^{-2}$  & 50  & 54  \\
  50 & 2.0 & 1   & $7.9\times10^{-3}$ & $2.2\times10^{-1}$  & 46  & 46  \\
  50 & 2.0 & 10  & $2.6\times10^{-2}$ & $1.8\times10^{-1}$  & 45  & 47  \\
  50 & 2.0 & 100 & $3.8\times10^{-2}$ & $7.1\times10^{-2}$  & 39  & 42  \\
\end{tabular}
\end{ruledtabular}
\end{table}

\begin{figure}[t]
\captionsetup{justification=raggedright, singlelinecheck=false}
  \includegraphics[width=0.48\textwidth]{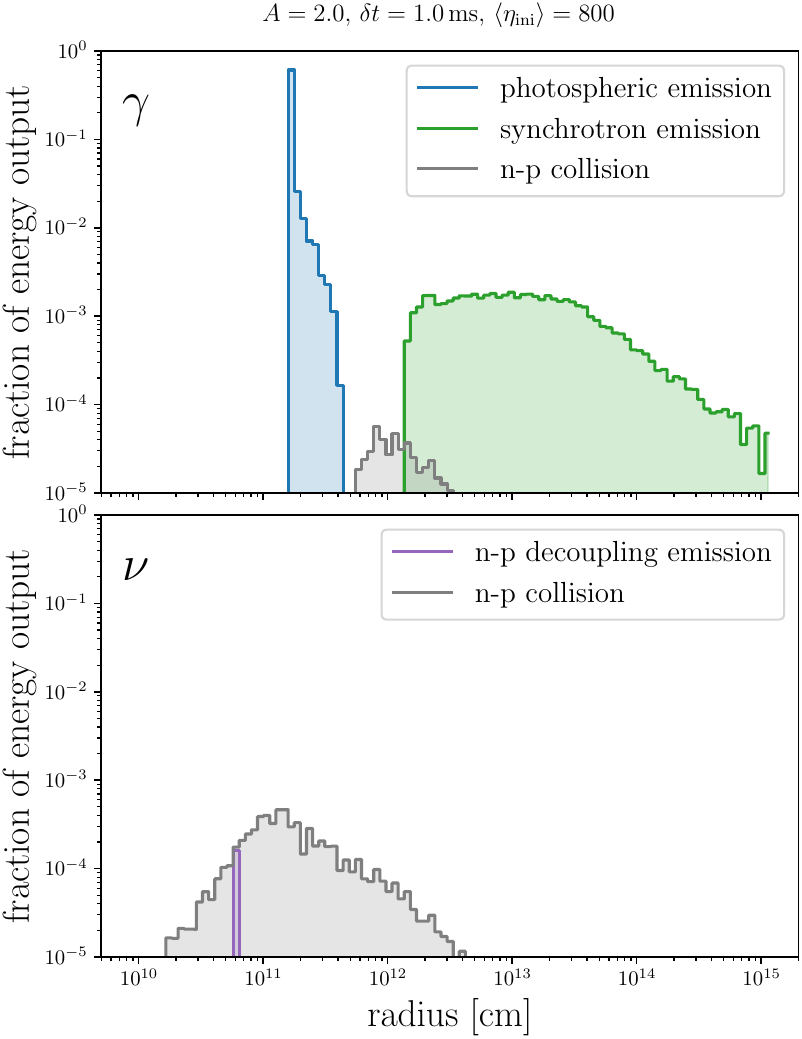}
  \caption{Same as Fig.~\ref{fig:FiducialEneOut}, but for the case  \( A = 2 \), \( \delta t = 1\,\mathrm{ms} \), and \( \left< \eta_\mathrm{ini} \right> = 800 \), that yields one of the most efficient photon emission.
  }
  \label{fig:MostPhoEneOut}
\end{figure}

\begin{figure}[t]
\captionsetup{justification=raggedright, singlelinecheck=false}
  \includegraphics[width=0.48\textwidth]{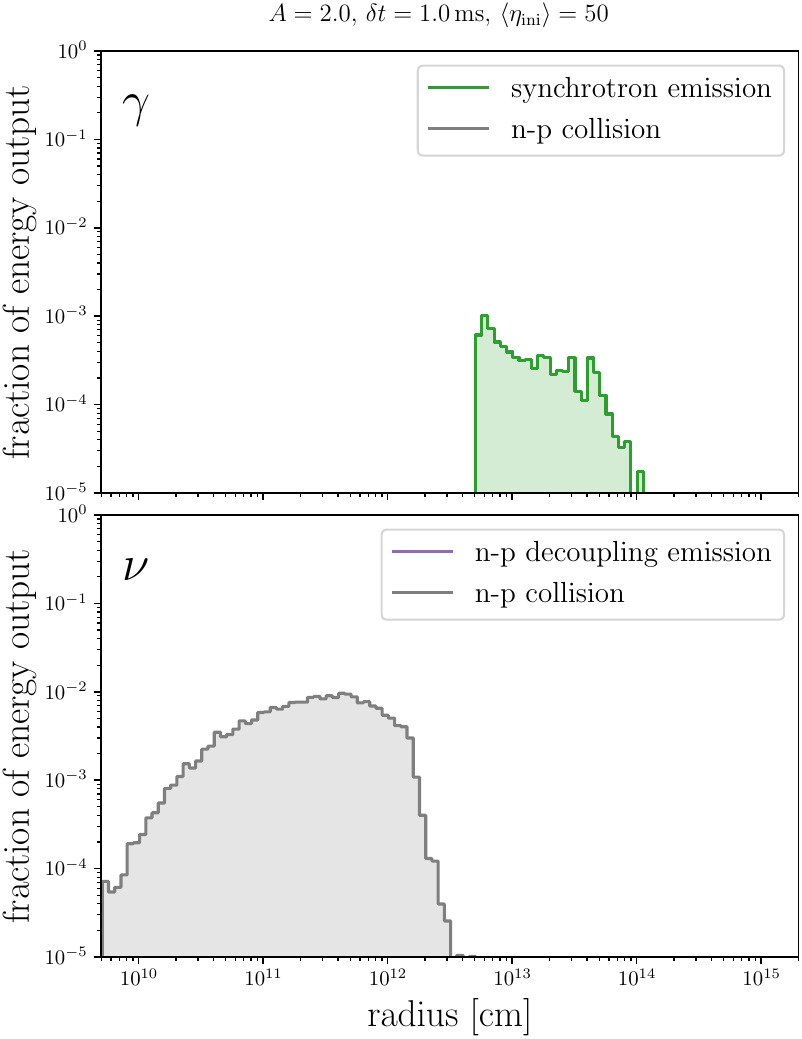}
  \caption{Same as Fig.~\ref{fig:FiducialEneOut}, but the case with \( A = 2 \), \( \delta t = 1\,\mathrm{ms} \), and \( \left< \eta_\mathrm{ini} \right> = 50 \), that yields the most efficient neutrino emission.
  }
  \label{fig:MostNeuEneOut}
\end{figure}

We perform simulations by varying the jet parameters: the amplitude of the variability \( A =0.5,\, 1,\,2\); variability timescale \( \delta t = 1\,\mathrm{ms},\,10\,\mathrm{ms},\,100\,\mathrm{ms} \); and the average baryon loading factor \( \left< \eta_\mathrm{ini} \right> = 50,\,200,\,800 \).
The isotropic jet luminosity initially injected is fixed at a relatively high value among the observed isotropic gamma-ray luminosities, \( L_\mathrm{j,iso} =3.5 \times 10^{52}\,\mathrm{erg\,s^{-1}}\), and shells are injected for a duration of \( \Delta t = 20\,\mathrm{s} \).
In total, we consider $3 \times 3 \times 3  = 27$ parameter sets.
The case with \( \left< \eta_\mathrm{ini} \right> = 200 \), \( A = 1 \), \( \delta t = 10\,\mathrm{ms} \) is adopted as the fiducial model.
A summary of the results for all parameter sets is presented in Table~\ref{tab:energy-results}.

\subsection{Energy dissipation in jets across the photosphere}
Figure~\ref{fig:FiducialEneOut} shows the energy dissipation processes and their dependence on radius in the fiducial jet.  
The upper and lower panels correspond to photon and neutrino emission, respectively.  
The vertical axis indicates the fraction of energy escaping from the system within each radial bin, normalized by the total injected energy.  
The first significant event in the innermost region is n–p decoupling.  
The accelerating shells undergo decoupling at a radius of \( r_\mathrm{np} \sim 9.4 \times 10^{10}~\mathrm{cm} \) (Eq.~\ref{r_np}) with a Lorentz factor of \( \Gamma_\mathrm{np} \sim 390 \) (Eq.~\ref{Gamma_np}).  
At larger radii, some shells emit photospheric radiation at \( r_\mathrm{ph} \sim 2.5 \times 10^{11}~\mathrm{cm} \) (Eq.~\ref{r_ph}) with a Lorentz factor of \( \Gamma_\mathrm{ph} \sim 1100 \) (Eq.~\ref{Gamma_ph}).  
We note that in our fiducial case with \( \langle \eta_\mathrm{ini} \rangle = 200 \), most of the injected shells experience neither n–p decoupling nor photospheric emission.  
Only those shells with sufficiently large initial baryon-loading factors, owing to the variability, and that can reach the corresponding critical radii without significant deceleration by internal shocks are able to produce these emissions.  
Nevertheless, the photospheric emission remains the dominant energy-loss process of the jet, accounting for about \( \sim 10\% \) of the injected jet energy, simply because of its high radiation efficiency.  
From Eqs.~(\ref{eq:E_decNu}) and (\ref{eq:Ephgam}), the energy loss via n–p decoupling is estimated to be very roughly \( \tau_\mathrm{dec}\, \kappa_\mathrm{dec}\, \Gamma_\mathrm{np} / \Gamma_\mathrm{ph} \sim 10^{-3} \) of that due to photospheric emission.

In addition to n–p decoupling and photospheric emission, shell-collision-induced emission occurs at $\gtrsim 2c\delta t \Gamma_\mathrm{ini}^2 \sim 6\times10^{10}\,\mathrm{cm}$.
Since efficient neutrino production requires $\tau_{\mathrm{np}}\gtrsim 1$ and a relatively large relative Lorentz factor, the neutrino emission induced by shell collisions peaks when $\tau_{\mathrm{np}}\sim 1$ (i.e., at $r 
\sim 2.3\times10^{12}\,\mathrm{cm}$).
At radii smaller than $\sim 9.2\times10^{12}\,\mathrm{cm}$, $\tau_\mathrm{T}>1$ and thus the shock is radiation mediated while at radii larger than $\sim 9.2\times10^{12}\,\mathrm{cm}$, shocks become collisionless and produce synchrotron emission as shown in Figure~\ref{fig:FiducialEneOut}. 
Evaluating at this transition radius,
the ratio of the energy output from n–p–collision neutrino emission to that from synchrotron emission is 
approximately $\kappa \tau_\mathrm{np} / \epsilon_\mathrm{e}\sim0.02$, 
where $\tau_\mathrm{np} = \sigma_\mathrm{np}/\sigma_\mathrm{T} \sim 1/20$, 
$\epsilon_\mathrm{e} \sim 1/3$, and $\kappa \sim 0.1$.
Significant shell-collision–induced emission primarily arises during the dissipation of the kinetic energy of the initially fastest shells.
In the fiducial case, the Lorentz factor of these initially fastest shells is set by $\sim\Gamma_{\mathrm{ph}}$.
The radius at which their kinetic energy is dissipated can then be estimated as
\[
\lesssim c\Delta t / N_{\mathrm{ph}}\, \langle \eta_{\mathrm{ini}} \rangle^{2}
   \sim c\,2\delta t/0.036\,\langle \eta_{\mathrm{ini}} \rangle^{2}
   \sim 6.6\times10^{14}\,\mathrm{cm}.
\]
Here, $N_{\mathrm{ph}}\sim 0.036N$ denotes the number of shells that have emitted photospheric radiation.
Figure~\ref{fig:FiducialEneOut} shows that significant shell-collision–induced emission occurs up to within this radius.

Shell collisions act as a relaxation process and the Lorentz-factor distribution becomes progressively concentrated around the center-of-momentum Lorentz factor.
This relaxation proceeds sequentially;
as the jet evolves, slower shells are successively overtaken by faster ones. 
Because each collision reduces the Lorentz factor of
the trailing shell, subsequent collisions propagate backward
through the flow, gradually driving the jet toward a homologous expansion.
At $\sim 4\times10^{4}\,\mathrm{s}$ in the central engine rest frame, which corresponds to the time when we terminated our calculation or $r\sim10^{15}\,\mathrm{cm}$ in Figure~\ref{fig:FiducialEneOut}, the average Lorentz factor of the shells is almost equal to the center-of-momentum Lorentz factor of $\sim 160$.
With this parameter, the jet dissipates mainly with photons not with neutrinos.

Figure ~\ref{fig:MostPhoEneOut} shows the energy dissipation processes and their dependence on radius in the case  \( A = 2 \), \( \delta t = 1\,\mathrm{ms} \), and \( \left< \eta_\mathrm{ini} \right> = 800 \), that yields one of the most efficient photon emission.  
The dynamics resembles the fiducial case qualitatively.
However, because $\delta t=1\,\mathrm{ms}$ is 10 times smaller than the fiducial case, 
shell-collision-induced emission start much inner than the np-decoupling radius and the photospheric emission radius.
When an accelerating shell experiences a collision as the faster shell, its Lorentz factor first drops abruptly due to the collision and then re-accelerates. 
As a result, the acceleration proceeds more slowly than the scaling expected for a single isolated shell. 
This slower acceleration also delays the decrease in density measured in the comoving frame, pushing the photospheric emission radius outward. 
Figure~\ref{fig:MostPhoEneOut} shows components corresponding to such delayed photospheric emission.

As in the fiducial case, the shell-collision–induced neutrino emission peaks at $\tau_{\mathrm{np}}\sim 1$, 
whereas snchrotron emission begins once $\tau_{\mathrm{T}}<1$, i.e., at $r\sim 1.5\times10^{12}\,\mathrm{cm}$.
The shell-collision–induced emission occurs within 
\[
\lesssim c\Delta t / N_{\mathrm{ph}}\, \langle \eta_{\mathrm{ini}} \rangle^{2}
   \sim c\,2\delta t/0.25\,\langle \eta_{\mathrm{ini}} \rangle^{2}
   \sim 1.4\times10^{14}\,\mathrm{cm}.
\]
At $r \sim 1.5\times10^{12}\,\mathrm{cm}$, where the shocks become collisionless, the system has already undergone substantial relaxation 
(i.e., the difference of the average Lorentz factor and the center-of-momentum Lorentz factor is much less than the initial average baryon loading factor $\langle \eta_\mathrm{ini}\rangle$). 
As a result, the radiation efficiency through synchrotron emission is lower than that obtained by \cite{Beloborodov2000} for the same variability amplitude ($A = 2$). In their case, most of the free energy supplied initially, $(\langle \eta_\mathrm{ini} \rangle - \Gamma_\mathrm{CM})\,N\,(m_\mathrm{n}^{(i)} + m_\mathrm{p}^{(i)})\,c^2$, is radiated away through shock-induced synchrotron emission.
 
The trend that photospheric emission becomes highly efficient, making photons the dominant radiative component, 
holds for sufficiently high average initial internal energy and fluctuation amplitude (i.e., the cases with $\langle \eta_{\mathrm{ini}} \rangle = 800$ or $\langle \eta_{\mathrm{ini}} \rangle = 200$ and $A \gtrsim 1$), 
because in these cases a substantial fraction of the shells fails to fully convert their internal energy into bulk kinetic energy before reaching the photosphere, leading to prominent photospheric emission.

On the other hand, for $\langle \eta_\mathrm{ini} \rangle = 50$, shells with large initial baryon-loading factors are absent, so efficient neutrino emission from n–p decoupling or photospheric emission does not occur, and energy dissipation takes place only through shell collisions.
Figure~\ref{fig:MostNeuEneOut} shows the energy dissipation processes and their dependence on radius in the case  
$\langle \eta_\mathrm{ini} \rangle = 50$, $A=2$, and $\delta t = 1\,\mathrm{ms}$,
that yields the most efficient neutrino emission.
Compared with Figure~\ref{fig:FiducialEneOut} and \ref{fig:MostPhoEneOut}, the larger number of shells with small Lorentz factors 
leads to frequent collisions at smaller radii, so that significant neutrino dissipation begins earlier 
and constitutes a larger fraction of the total energy budget. 
Since the jet has already relaxed in the region with $\tau_\mathrm{T}>1$, 
synchrotron emission is inefficient. 
Consequently, under this parameter set the jet is bright in neutrinos but dim in photons.

\begin{figure}[t]
    \captionsetup{justification=raggedright, singlelinecheck=false}
    \includegraphics[width=0.49\textwidth]
    {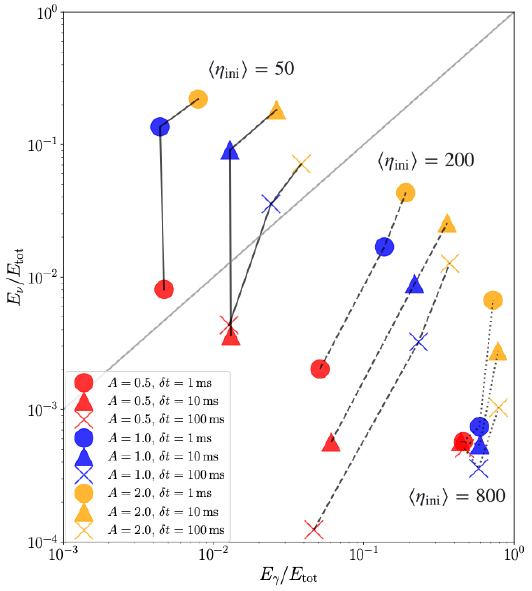}
  \caption{Photon radiation efficiency versus neutrino radiation efficiency, and their dependence on the properties of a time-variable, neutron-loaded jet.
  The isotropic luminosity and duration of the jet are fixed to be $L_\mathrm{j, iso} = 3.5\times 10^{52}\,\mathrm{erg\,s^{-1}}$, and $\Delta t = 20\,\mathrm{s}$ respectively.}
  \label{fig:efficiency_all}
\end{figure}

Figure~\ref{fig:efficiency_all} shows a summary of the fractions of the energy output.
The horizontal axis gives the total energy dissipated as photons divided by the injected energy,
and the vertical axis gives the total energy dissipated as neutrinos divided by the injected energy.
As discussed above, the locations of the points in Figure~\ref{fig:efficiency_all} 
reflect the dominant dissipation channels for the corresponding system parameters.
A larger amplitude of the time variability, characterized by a larger value of $A$, enhances more violent shell collisions (i.e., those with large relative Lorentz factors) during relaxation, thereby increasing the overall efficiency of energy dissipation.

A larger amount of the baryon loading, characterized by a smaller value of $\langle \eta_{\mathrm{ini}} \rangle$ and/or the smaller variability time scale $\delta t$ shift the dissipation inward, leading to more efficient neutrino emission and a corresponding reduction in photon production.
For $\langle \eta_\mathrm{ini} \rangle \lesssim 50$, $\delta t = 1\,\mathrm{ms}$, and $A \gtrsim 1$, the neutrino emission becomes efficient ($\gtrsim 10\%$), whereas the photon efficiency remains below $1\%$.
Although we did not perform a quantitative spectral analysis, the Yonetoku~\cite{Yonetoku+2004} and Amati relations~\cite{Amati2006} suggest that these neutrino-bright, photon-faint population identified in this work would have relatively soft spectral peaks.
Conversely, larger $\langle \eta_\mathrm{ini} \rangle$ and/or $\delta t$ lead outer collisions, reducing neutrino emission, and enhancing photon emission from collisionless shocks.
For larger values of $\langle \eta_\mathrm{ini} \rangle$, photospheric emission becomes increasingly dominant, leading to higher photon efficiency.
For $\langle \eta_\mathrm{ini} \rangle \gtrsim 800,\,$ or 
$\langle \eta_\mathrm{ini} \rangle \gtrsim 200$ and $A \gtrsim 1$, the gamma-ray emission is efficient, as in typical GRBs, while the radiative efficiency of GeV–TeV neutrinos remains at the level of $0.1$-$10\%$.

\begin{figure}[htbp]
  \centering
  \captionsetup{justification=raggedright, singlelinecheck=false}
    \includegraphics[width=0.49\textwidth]
     {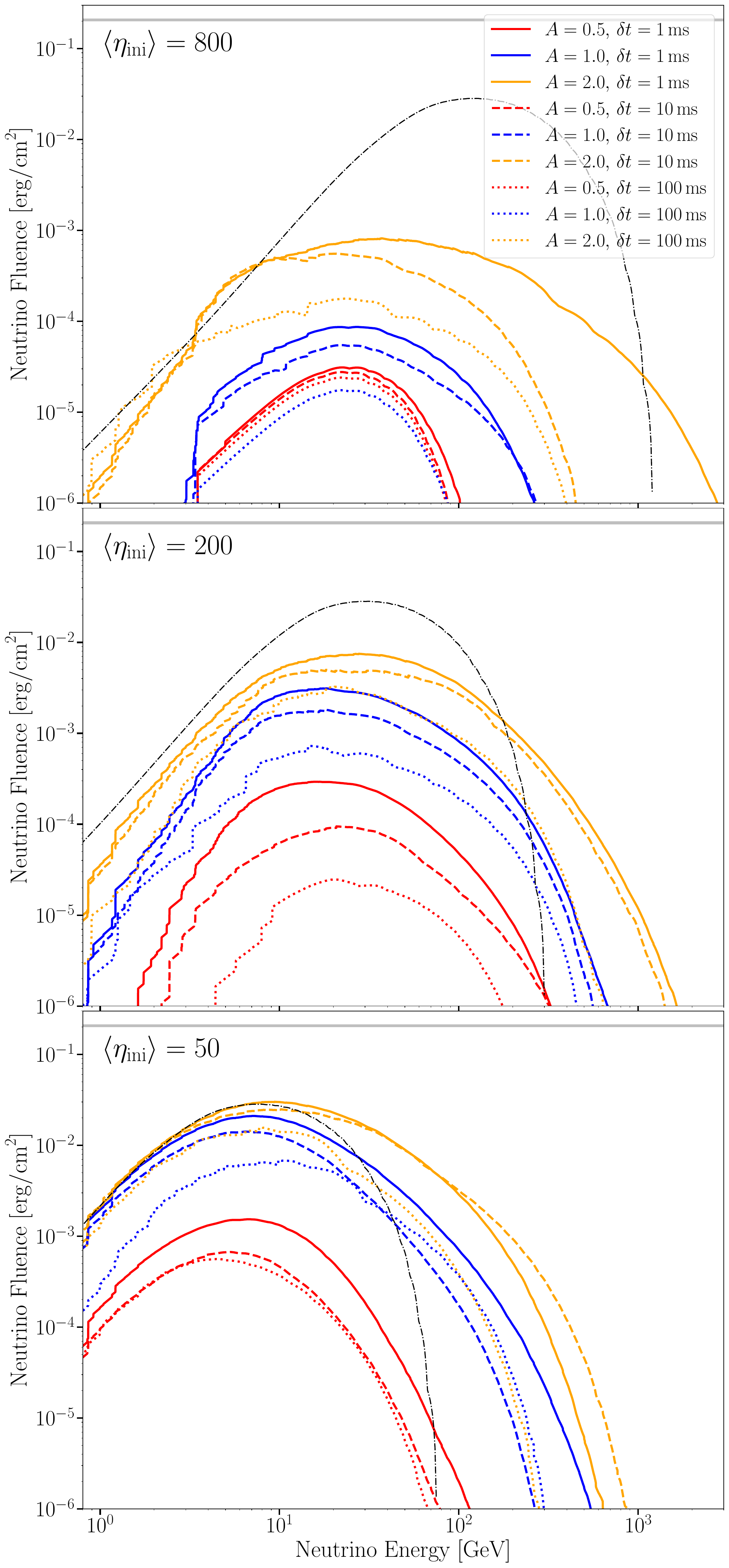}
  \caption{GeV-TeV neutrino counterparts of time-variable neutron-loaded jets with various statistical inhomogeneity. 
  The isotropic luminosity and duration of the jet, and the luminosity distance are fixed to be $L_\mathrm{j, iso} = 3.5\times 10^{52}\,\mathrm{erg\,s^{-1}}$, $\Delta t = 20\,\mathrm{s}$, and $d_\mathrm{L} = 100\,\mathrm{Mpc}$, respectively.
  The black dotted lines represent the IceCube templates for neutrinos from collision scenario, where the relative Lorentz factor is fixed to $\Gamma_\mathrm{rel}\sim2$~\cite{Murase2022,IceCubeCollab2023}. 
  }
  \label{fig:NeuEneSpe}
\end{figure}

\subsection{GeV-TeV neutrino counterparts}

Figures~\ref{fig:NeuEneSpe} show the neutrino energy spectra originated from inelastic neutron-proton interactions with assuming a luminosity distance of \(d_\mathrm{L} = 100\,\mathrm{Mpc}\) for all the cases.
Each curve represents the neutrinos produced during the dynamics from n–p decoupling and those produced from shell collisions for each parameter set.
For comparison, the IceCube template~\cite{Murase2022} is plotted as a dotted-dash line.
This template~\cite{Murase2022} provides the neutrino energy spectrum 
for collisions with a relative Lorentz factor of $\sim 2$, given an input bulk Lorentz factor. 
In our comparison, we set the bulk Lorentz factor to $\langle \eta_\mathrm{ini}\rangle$, and the normalization of this template, $\xi_\mathrm{N} E_\mathrm{\gamma, iso}$, which corresponds to the kinetic energy of the interacting flow that experienced an n-p collision with a relative Lorentz factor of $\sim 2$, to $E_{\mathrm{iso}}= L_\mathrm{j,iso} \Delta t =7\times 10^{53}\,\mathrm{erg}$~\cite{Murase2022}.

The above-mentioned normalization of the IceCube template corresponds to the fluence that would be obtained if all of the injected kinetic energy underwent an n-p collision with a relative Lorentz factor of $\sim 2$.
Only the cases with sufficiently large baryon loading, large fluctuation amplitude, and short variability timescale (i.e., $\langle \eta_{\mathrm{ini}} \rangle = 50$, $A = 4$, and $\delta t \lesssim 10\,\mathrm{ms}$) yield fluences comparable to this normalized IceCube template, 
whereas cases with other parameter choices produce lower fluences.
Regarding the relative fluence among the parameter sets examined in this study,  
large values of $A$ and small values of $\langle \eta_{\mathrm{ini}} \rangle$ and/or $\delta t$ lead to both higher neutrino-production efficiency and a larger total neutrino fluence, as Figure~\ref{fig:efficiency_all} also shows.

The peak neutrino energy in the IceCube template is expressed as
$
E_\nu \sim 0.1\,\Gamma \Gamma'_\mathrm{rel} m_\mathrm{u}c^2, 
\quad \Gamma \sim \langle \eta_\mathrm{ini}\rangle, 
\quad \Gamma'_\mathrm{rel} \sim 2
$~\cite{Murase2022}.
While the IceCube original template assumes a single shell Lorentz factor and a single relative velocity,
actual jets involve a diversity of collisions, and the resulting spectra are therefore generally broader.
We find that the peak energy of the GeV-TeV neutrinos lies in $10$-$30\,\mathrm{GeV}$, 
with the high-energy tail extending into the TeV range not primarily due to shorter variability timescales $\delta t$, 
but instead when the amplitude of the time variability $A$ is large.
Neutrino production is mainly from n-p decoupling for parameter sets with high $\langle \eta_\mathrm{ini} \rangle$ and small variability, $\langle \eta_\mathrm{ini} \rangle = 800$, $A = 0.5$ with $\delta t = 1, 10, 100\,\mathrm{ms}$, and $A = 1$ with $\delta t = 100\,\mathrm{ms}$. 
In these cases, the peak energy in the GeV-TeV range is consistent with the expected neutrino energy from n-p decoupling, estimated to be $\sim 20\,\mathrm{GeV}$ (Eq.~(\ref{eq:E_decNu}) and \cite{BahcallMeszaros2000, Murase2022}). 
In contrast, neutrino emission is dominated by shell collisions for all other parameter sets, where the typical neutrino energy is determined by the product of the nucleon energy of the fast shell and the fraction of energy a neutrino acquires per scattering 
(i.e., $E_\nu \sim 0.1 \Gamma_\mathrm{r} m_\mathrm{u} c^2$; Eq.~(\ref{eq:Enu_eachShell}), \cite{Murase2022}), 
and the peak energy also lies around $\sim 10$-$30\,\mathrm{GeV}$.

For moderate initial baryon-loading factors, \(\langle \eta_\mathrm{ini} \rangle \lesssim 200\), the resulting neutrino spectra are broadly consistent with the IceCube template in the sense that the spectral peak is shifted toward higher energies in a similar manner.  
However, for a higher baryon-loading factor of \(\langle \eta_\mathrm{ini} \rangle = 800\), this shift in the peak energy becomes suppressed.  
In this high-\(\eta\) regime, the dominant dissipation channel for neutrinos depends on the amplitude and timescale of the fluctuations:  
for relatively small fluctuations (\(A=0.5\) with \(\delta t = 1, 10, 100\,\mathrm{ms}\), and \(A=1\) with \(\delta t = 100\,\mathrm{ms}\)), neutrino production is mainly governed by n-p decoupling, whereas for larger fluctuations it is dominated by shell collisions.  
When \(\langle \eta_\mathrm{ini} \rangle = 800\) and the fluctuations are large, a significant number of shells initially have very high \(\eta_\mathrm{ini}\).  
However, such shells reach the photosphere while still accelerating, and their Lorentz factors are limited to 
\(\Gamma_\mathrm{ph} \sim 670\) (see Eq.~(\ref{Gamma_ph})).  
Consequently, in cases with both large fluctuations and high \(\langle \eta_\mathrm{ini} \rangle\), the maximum Lorentz factor within the system is capped, and the peak energy of the neutrino spectrum remains around \(\sim 10\,\mathrm{GeV}\).  

\begin{figure}[t]
\captionsetup{justification=raggedright, singlelinecheck=false}
\includegraphics[width=0.48\textwidth]{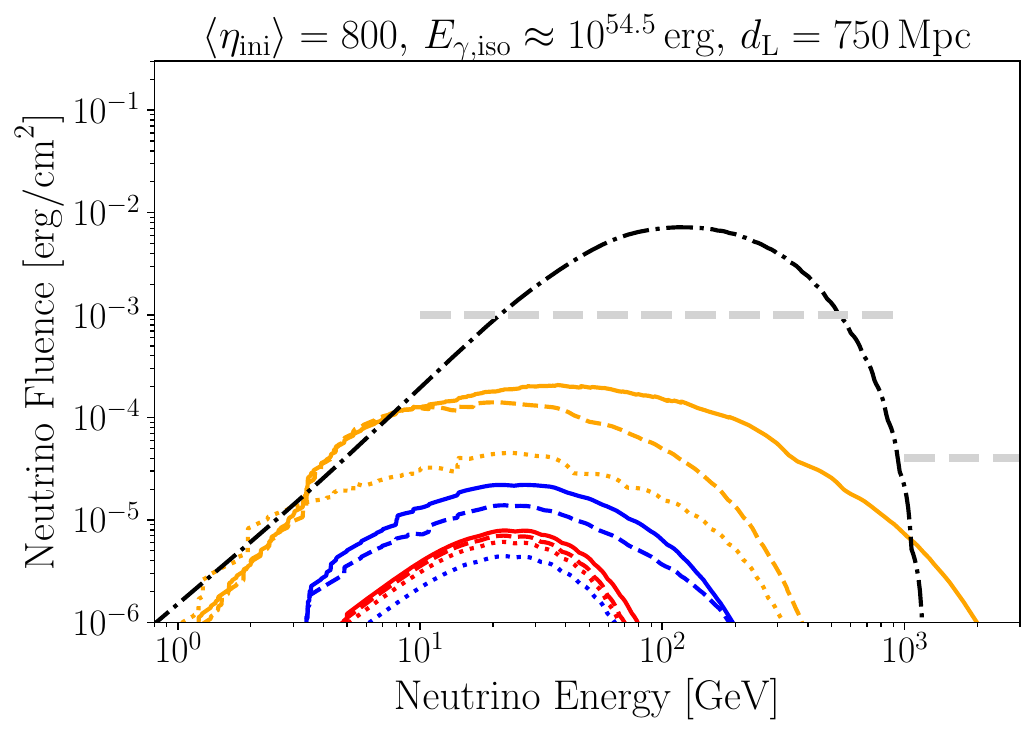}
  \caption{
  Same as the upper panel of Figure~\ref{fig:NeuEneSpe}, but with the fluence normalization modified to match that of the BOAT GRB ($E_\mathrm{iso}=\xi_\mathrm{N}E_\mathrm{\gamma, iso} = 10^{55}\,\mathrm{erg},\, d_\mathrm{L}=750\,\mathrm{Mpc}$).
  }
  \label{fig:BOATcomparison}
\end{figure}

The IceCube Collaboration~\cite{IceCubeCollab2023} discussed the properties of the jet based on the non-detection of neutrinos from GRB 221009A (the BOAT GRB), the brightest GRB ever observed, using the theoretical template~\cite{Murase2022}.
Figure~\ref{fig:BOATcomparison} shows the result 
from \( \left< \eta_\mathrm{ini} \right> = 800 \) in Figure~\ref{fig:NeuEneSpe} with the normalization for the BOAT GRB, i.e., we set the luminosity distance to \( d_\mathrm{L} = 750~\mathrm{Mpc} \) and just scale the result by a factor of \( 10^{55}~\mathrm{erg} / 7 \times 10^{53}~\mathrm{erg} \) to match the total energy of the BOAT~\cite{Frederiks+2023, Lesage+2023, Sato+2025}.
The thick light-gray dashed line shows the upper limit set by IceCube observations~\cite{IceCube2024_GeV-TeVserch}.

Figure~\ref{fig:BOATcomparison} indicates that none of the parameter sets we consider in this study for the BOAT GRB jets have yet been ruled out by the GeV-TeV neutrino observations. 
As demonstrated in this work, there exists an anti-correlation between the photon efficiency and the neutrino efficiency.
Therefore, jets such as that of the BOAT GRB are not neccessarily favorable targets for neutrino detection, even though previous studies implicitly expected that GRBs bright in photons would also be bright in GeV-TeV neutrinos.
In fact, the IceCube template has a normalization parameter $\xi_{\mathrm{N}}$ to be constrained observationally. 
(Here, $\xi_{\mathrm{N}} E_{\gamma,\mathrm{iso}}$ corresponds to the kinetic energy of the interacting flow that undergoes an n–p collision with a relative Lorentz factor of $\sim 2$.) 
According to our results, theBOAT GRB jet has $\xi_{\mathrm{N}} \lesssim 0.1$, which is not constrained by the observational data of the BOAT GRB, as shown in Fig.~\ref{fig:efficiency_all} \citep{Murase2022, IceCubeCollab2023}.

So far, searches for GeV–TeV neutrinos from GRBs also have been done for a stacking analysis of 2268 GRBs detected on 2012 April 26 and 2020 May 29~\cite{IceCube2024_GeV-TeVserch} as well as 
the BOAT GRB as an individual source.
The stacked 2268 GRBs include events with a wide range of luminosities and distances; based on event-rate estimates, most of them are likely at cosmological distances with typical luminosities~\cite{PalmerioDaigne2021}.

\begin{figure}[t]
\captionsetup{justification=raggedright, singlelinecheck=false}
\includegraphics[width=0.48\textwidth]{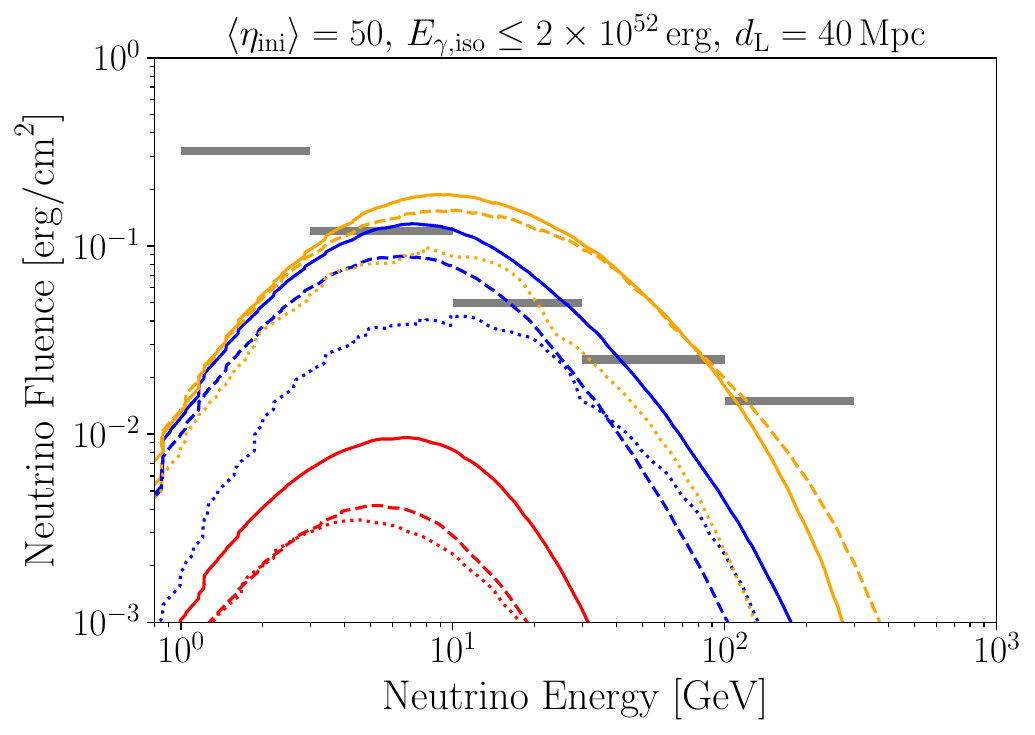}
  \caption{
   Same as the lower panel of Figure~\ref{fig:NeuEneSpe}, but with the fluence normalization modified to $E_\mathrm{iso}=L_\mathrm{j,iso}\Delta t = 7\times 10^{53}\,\mathrm{erg},\, d_\mathrm{L}=40\,\mathrm{Mpc}$.
  In these parameters, the expected observed photon energy remain at $E_{\gamma,\mathrm{iso}} \leq 2\times 10^{52}\,\mathrm{erg}$.
  The gray horizontal line represents the sensitivity that will be achieved with the IceCube upgrade~\cite{KobayashiYukiho2025}.}
  \label{fig:ForDetection}
\end{figure}

Our results indicate that, for neutrino detection, GRBs that are fainter in photons than the typical bright GRBs can be more promising. 
From the observation~\cite{Yonetoku+2004, Amati2006}, they would have relatively soft spectral peaks.
Therefore, GRBs that appear faint in photon observations, such as those detectable by instruments like the Einstein Probe~\cite{EinsteinProbe2015}, are expected to be favorable targets.
In Figure~\ref{fig:ForDetection}, we compare the result from \( \left< \eta_\mathrm{ini} \right> = 50 \) in Figure~\ref{fig:NeuEneSpe} with the luminosity distance set to \( d_\mathrm{L} = 40~\mathrm{Mpc} \) and the sensitivity of IceCube up-grade~\cite{KobayashiYukiho2025}.
The physical quantities plotted are the same as in Figure \ref{fig:NeuEneSpe}.
As shown in Figure~\ref{fig:ForDetection}, a jet with $\left< \eta_\mathrm{ini} \right> = 50$, which is expected to be observed as a soft GRB with $E_{\gamma,\mathrm{iso}} \leq 2\times 10^{52}\,\mathrm{erg}$, can be detected as a single source at distances of 40 Mpc when the fluctuations are large (the blue or yellow cases). 
In practice, stacking analyses of soft, faint GRBs located somewhat farther away would also be an effective strategy.

Finally, we comment on very high-energy (TeV--PeV) neutrinos within our model framework.
In this work, we did not consider cosmic-ray acceleration, in order to avoid additional uncertainties associated with particle acceleration.
Nevertheless, collisionless shocks can accelerate cosmic rays, which then interact with photons in the shocks through the \(p\gamma\) process to produce TeV--PeV neutrinos~\cite{WaxmanBahcall1997, MuraseNagataki2006}.
From the non-detection of very high-energy neutrinos from the BOAT GRB, one can place the strongest constraint on the product of the cosmic-ray loading factor \(\xi_\mathrm{CR}\equiv {\cal E}_\mathrm{CR,iso}/{\cal E}_\mathrm{\gamma,iso}\) and the optical depth of the \(p\gamma\) process \(f_{p\gamma}\);
\[
\mathrm{min}\!\left[\frac{\xi_\mathrm{CR}}{0.1},\, \frac{f_{p\gamma}}{0.01}\times\frac{\xi_\mathrm{CR}}{10}\right]
\lesssim 2\times \left(\frac{{\cal E}_\mathrm{\gamma,iso}}{10^{54.5}\,\mathrm{erg}}\right)^{-1},
\]
as discussed in~\cite{Murase2022}.
Here, \({\cal E}_\mathrm{CR,iso}\) denotes the isotropic-equivalent energy carried by the accelerated cosmic rays, and \({\cal E}_\mathrm{\gamma,iso}\)  corresponds to the synchrotron radiation energy generated in internal shocks.
In the cases corresponding to the BOAT GRB (\(\langle \eta_\mathrm{ini}\rangle = 800\)), our results indicate \({\cal E}_\mathrm{\gamma,iso} \lesssim 0.01\,E_\mathrm{iso}\).
Since collisionless shocks occur at radii \(r \gtrsim 10^{12-13}\,\mathrm{cm}\) for the cases with \(\langle \eta_\mathrm{ini}\rangle = 800\), the value of \(f_{p\gamma}\) can vary significantly~\cite{Murase2022}.
In the most optimistic case for the constraint on the acceleration efficiency, we obtain a constraint,
\[
\xi_\mathrm{CR} \lesssim 2\times100\times0.1 = 20.
\]
This limit is weaker than that inferred by assuming that the observed gamma rays are entirely produced via synchrotron radiation from collisionless shocks (i.e., \({\cal E}_\mathrm{\gamma,iso}/10^{54.5}\,\mathrm{erg}\approx 1\))~\cite{Murase2022}, 
because in our model the efficient gamma-ray emission is predominantly due to photospheric emission.

\section{Summary}\label{sec:summaryANDdiscussion}
We develope a shell-based model of time-variable, neutron-loaded jets, incorporating sub-photospheric dissipation, and follow its relaxation.
We have shown that the statistical variability imprinted at the stellar breakout of collapsar jets governs the diversity of their dissipation histories both below and above the photosphere.
The larger amplitude of the time variability (in this study which is represented by larger $A$) generally leads to more frequent violent collisions 
(i.e., those with large relative Lorentz factors) to relax.
When the initial specific energy (in this study which is represented by $\langle \eta_\mathrm{ini} \rangle$) and/or the variability time scale (in this study which is represented by $\delta t$) are large, 
collisions typically occur at relatively large radii, resulting in reduced neutrino efficiency 
and enhanced photon efficiency, 
while the dominant contribution to the dissipated photon energy comes from highly efficient photospheric emission.
On the other hand, when $\langle \eta_\mathrm{ini} \rangle$ and/or $\delta t$ are small, 
collisions typically occur at deeper inside resulting in enhance neutrino efficiency and reduced photon efficiency.
GeV–TeV neutrino emission peaks around $\sim10\,\mathrm{GeV}$, with a TeV tail for strongly variable jets. 
For jet parameters that realize high photon radiative efficiency (i.e., those corresponding to typical GRBs), the neutrino efficiency remains below 10\%. In contrast, in cases with strong variability and large baryon loading(i.e., low initial specific energy), it can reach $\sim20$\% while the photon emission efficiency stays below 1\%. Such jets may correspond to gamma-ray–dim but neutrino-bright GRBs, making them promising targets for future neutrino observations.

\begin{acknowledgments}
We are very grateful to Wataru Ishizaki, Yukiho Kobayashi, and Kohta Murase for many fruitful discussions and advices.
N.K. is  very grateful to Naoki Otani and Sei Ieki for their kind instruction on GEANT4.
 This work is supported by AGS RISE Program at Tohoku University (N.K.), and JSPS KAKENHI grant Nos. JP24K00668, JP23H04899, JP22H00130 (K.K.).
\end{acknowledgments}

\bibliographystyle{apsrev4-2}  
\bibliography{apssamp}

\end{document}